\documentclass{kapproc} 

\usepackage{procps} 

\usepackage[dvips]{graphicx}

\upperandlowercase

\kluwerbib 

\draft

\begin{document}

\articletitle
[Bars, Spirals and Warps Result from the Same Fundamental Causes]
{A Unified Picture of Disk Galaxies where Bars, Spirals and Warps
Result from the Same Fundamental Causes}

\author{Daniel Pfenniger and Yves Revaz} 
 
\affil{Geneva Observatory, University of Geneva,
  CH-1290 Sauverny, Switzerland}

\begin{abstract}
  Bars and spiral arms have played an important role as constraints on
  the dynamics and on the distribution of dark matter in the optical
  parts of disk galaxies.  Dynamics linked to the dissipative nature
  of gas, and its transformation into stars provide clues that spiral
  galaxies are driven by dissipation close to a state of
  \textit{marginal stability} with respect to the dynamics in the
  galaxy plane.  Here we present numerical evidences that warps play a
  similar role but in the transverse direction.  N-body simulations
  show that typical galactic disks are also marginally stable with
  respect to a bending instability leading to typical observed warps.
  The frequent occurrence of warps and asymmetries in the outer
  galactic disks give therefore, like bars in the inner disks, new
  constraints on the dark matter, but this time in the outer disks.
  If disks are marginally stable with respect to bending
  instabilities, our models suggest that the mass within the HI disks
  must be a multiple of the detected HI and stars, i.e., disks must be
  heavier than seen.  But the models do not rule out a traditional
  thick halo with a mass within the HI disk radius similar to the
  total disk mass.
\end{abstract}


\section{Introduction} 

Over the years we discover progressively the true nature of galaxies
by confronting observations to theory and vice versa.  A simple 
tool to rank the importance of  the various physical ingredients at
play in galaxies is a more complete form of the virial
theorem than usually discussed:
\begin{equation}
{\scriptstyle{\frac{1}{2}}} \ddot I ~=~ 
  \underbrace{2 E_{\rm kin}}_{>0} ~+~ 
  \underbrace{E_{\rm grav}}_{<0} ~+~ 
  \underbrace{3 P_{\rm int} V}_{>0}
  ~\underbrace{-~3P_{\rm ext}V}_{<0} ~+~ 
  \underbrace{E_{\rm mag}}_{>0} 
  \ldots ~\approx~ 0\ . 
\end{equation}
It allows to rank the main energies (bulk kinetic, gravitational,
internal and external pressures, magnetic, etc.)  determining the
system equilibrium measured by its moment of inertia $I$ in the volume
$V$.  Clearly the \textit{magnitudes of the interacting energies}
(mainly bulk kinetic against gravitational energy in galaxies) rank
the importance of each factor.  A Taylor expansion of this equation in
time also shows that an evolution along a sequence of quasi-steady
equilibria is determined to first order by the \textit{magnitude of
  the interacting powers} (mainly gas cooling against mechanical
heating power).

The main point emphasized here is that the same rules
found to well explain to first order the horizontal properties of
spiral galaxies, i.e., gravitational physics supplemented by energy
dissipation, are also able to explain the ubiquitous warps.  By
comparing numerical simulations of thin self-gravitating disks to the
observed properties of spirals, in particular to their frequent warps,
a coherent dynamical and evolutive picture of spiral galaxies emerges,
with the suggestive hint that two types of dark matter are involved:
1) the classical extended dark halos much thicker than the disks, and
2) a dark component coeval with the HI disks similar to the one
proposed in Pfenniger \& Combes (1994).

Nowadays the need of at least two types of dark matter is actually
well motivated.  From the big-bang predicted baryogenesis most
of the baryons remain to be found, and on the other hand, also from
cosmology non-baryonic dark matter is required to
obtain a coherent description of large scale structure formation and
the Universe cosmological parameters.  Since baryons are known to be
strongly dissipative and sometimes collisional, contrary to the
expected non-baryonic dark matter, we have no grounds to expect that
in galaxies their respective spatial distributions should coincide.

\section{The role of bars and spirals}
The bar instability has been used in the 70's for supporting the idea
that an extended dark matter halo must exist to prevent bar formation
(Ostriker \& Peebles, 1973).  Indeed the early N-body simulations
showed that bars result spontaneously from a dynamical plane
instability in a collisionless disk with an initial flat core.  But
because theoreticians were wrongly assuming that bars were exceptional,
they imagined a hot and massive collisionless component coexisting in
the optical disks as a solution to prevent the quick formation of
bars, despite the awareness by skilled observers such as de Vaucouleurs 
that bars are frequent.  Subsequent higher
resolution and infrared observations revealed that bars are in fact
even more frequent and found in a majority of spirals, as reminded 
in several papers at this conference.

The reverse problem was thus discussed many times: how bars and hot
dark halos can coexist, since dark halos were then no longer viewed as
hypothetical.  It was also understood over the years that barless
disks can exist when the central density profile is too centrally 
concentrated.

Spiral density waves are a more general but less robust version of the
bar phenomenon.  The main reason for spiral formation is now well
understood as resulting from the non-linear growth of a 
spontaneous gravitational instability in the disk
plane with the same origin as for bars: a kinematically too cold disk
is gravitationally unstable, and the non-linear result is typically a
bar in the initial flat core and spiral arms in the outer
differentially rotating region.
Without invoking more than Newtonian physics it was also found that
the typical double exponential disks profiles are also a natural
asymptotic state for a collisionless disk passing through a bar
instability (Pfenniger \& Friedli, 1991).

The non-linear structures resulting from the gravitational instability
are never strictly stationary, but evolve secularly (over several
rotational periods).  In pure collisionless disks they tend to destroy
the spiral arms and later bars, so obviously something must regenerate
them in real galaxies.

The theory of bars and spirals is presently incomplete because the
full self-consistent problem is very non-linear.  Thus no 
analytical theory able yet to \textit{predict} the full development of
strongly rotating collisionless self-gravitating disks.  Only the
brute force N-body simulations are able to do it.  

Since the Newtonian physics involved in these N-body experiments is
very well understood, and that the numerical codes
can to a large extent be trusted because different versions 
implementing different
techniques have been developed and checked over several decades by
many groups, we can use these N-body techniques to predict
and explain the behaviours of galaxies.  The results of such
N-body simulations should be taken as seriously as analytical
developments in celestial mechanics.  As example of success of N-body
techniques is the prediction that bars may evolve into peanut-shaped
structures.  This was first empirically found in N-body experiments
(Combes \& Sanders, 1981; Combes et al., 1990), understood
theoretically (Pfenniger \& Friedli, 1991), and later confirmed by 
observations (e.g., Bureau \& Freeman, 1999).

Coupled to this well understood underlying physics, numerous
independent studies of the mass to light ratio in the optical parts of
spirals have determined that a substantial fraction of the gravitating
mass there is well explained by the detected baryons (e.g. Sancisi,
2003).  This ensures that we have at least a basic physical
understanding of the inner parts of galaxies.
 
\section{The role of gas}
Since disk galaxies as star producing systems must contain also a lot
of gas, its effects must be considered on the long run.  First, dust
polluted gas is very efficient to lose its thermal energy by infrared
radiation, so galaxies must be seen, besides in first approximation as
rotating self-gravitating objects, in second approximation also as
energy dissipating structures.  Gravitationally bound rotating
structures slowly losing energy tend to rapidly converge towards thin
disks because then angular momentum is a quantity much harder to
dissipate away than energy.

So the frequent occurrence of spiral arms (after all disk galaxies are
called spirals!)  follows directly from the constant competition
between the effect of cooling driving disks towards the gravitational
Safronov-Toomre instability threshold.  Any further cooling leads to
strong reaction from the disk by dynamical heating.  As long as gas
cooling continues to be efficient the natural long term state of
spiral galaxies is therefore to stay close to the marginal stability
threshold.

Numerous studies show that galactic disks, including the Milky Way,
have disks close to a marginal stability state with a Safronov-Toomre
parameter $Q$ close to unity.  Because of this marginal state, spiral
galaxies do react strongly to other perturbations such as galaxy
interactions by amplifying spiral arms.

Many studies have been trying to determine whether spirals and bars
result either from galaxy interactions or from a proper disk
instability.  The more fundamental cause of spirals and bars is
actually the internal marginal stability state making galactic disks
very reactive to various perturbations.  Even small satellite
interactions trigger large responses from a disk in the form of grand
design spirals.  The name of ``spirals'' for disk galaxies is in the
end an excellent way to characterize their close to marginal stability
state, showing both that dissipation acts and dynamics reacts.

A corollary of such a marginally stable state is that a steady state
is unlikely.  Instead evolution is to be expected as long as the
marginal stability state is maintained by the competing factors, gas
cooling against dynamical heating.
 
As by-product the large scale dynamical instability of galactic disks
leads to local interstellar gas compression, shocks and turbulence,
cascading down to smaller scale gas instabilities (e.g., Fleck, 1981;
Elmegreen, 2004).  At the bottom of the cascade the most visible effect
of the gas ``turbulence'' is star formation (Klessen, 2004), which
implies gas consumption.  The long term effects of large scale
instabilities is to transform progressively the dissipative component
into a collisionless stellar component.  By consuming gas the cooling
agent becomes rarer, and by forming stars the dynamical heating more
effective in counter-balancing gas cooling.  In addition the
mechanical energy output produced especially by massive stars provides
a second important source of heating competing gas cooling.  

\section{Constraints on dark matter forms}
So the slow transformation of matter from gas rich, but also dark
matter rich disks to gas poor, star rich and dark matter poor
structures already indicates that the above picture is broadly
consistent.  Gas poor disk galaxies (S0's, Sa's) have namely typically
less prominent and open spiral arms in more symmetric disk, while gas
rich spirals (Sd's, Sc's) have large open spirals in irregular disks.
The fact that along the spiral sequence the visible gas represents
always a minor fraction of the mass indicates that some of the dark
mass must be gas in order to be able to form subsequently all the
stars that are seen in S0's and Sa's (Pfenniger et al., 1994).

However the fact that S0's and Sa's still contain a fraction of dark
mass while showing very little star formation indicates also that some
of the dark mass is in a form that cannot easily form stars.  
Around 40\% of the total mass within the HI disk radius might be in a
dark collisionless form.  Therefore the above considerations show
already that we can have a consistent dynamical picture of disk
galaxies including the gas and star formation aspects provided that
\textit{two} forms of dark matter exist: one, close to the visible
gaseous form for explaining the properties of the spiral sequences as
an evolutive sequence of dissipative gravitating disks, and one
non-gaseous form for explaining the remaining ``indestructible'' dark
mass in the evolved part of the sequence, the S0's and Sa's.

\section{The role of warps}

All these considerations have been made considering the plane dynamics
of spiral galaxies, except for the bulge growth via vertical
instabilities in the inner stellar disk.  

But what about the dynamical effects transverse to the disks in the
outer regions?  Namely, a notorious puzzle in spiral galaxies is the
ubiquitous warp phenomenon which has eluded a clear explanation up to
now.  For instance warps are unlikely to result from resonant normal
modes, because the soft edges of galaxy disks damp discrete modes
(Hunter \& Toomre, 1969).  Normal modes generated by massive inclined
dark halo (Dekel \& Shlosman, 1983; Sparke, 1984; Sparke \& Casertano,
1988) are ruled out by dynamical friction that damp the warp in a few
dynamical times (Dubinski \& Kuijken, 1995).  Only particular triaxial
halos can produce a torque that leads to a warp with a straight line
of node and negligible back reaction (Petrou, 1980).  Interactions are
efficient in warping disks (Hernquist, 1991; Huang \& Carlberg, 1997),
however they cannot be invoked in isolated warped galaxies.

Warps are especially obvious in the HI outer disks, but to a lesser
amplitude the stellar disks are also warped.  Statistics of warps in
HI (Bosma, 1991; Richter \& Sancisi, 1994; Garcia-Ruiz et al., 1998)
and in the optical band (Reshetnikov \& Combes, 1998, 1999; 
Sanchez-Saavedra et al., 1990; Sanchez-Saavedra et al., 2003) reveal that
more than half the spiral galaxies are warped and asymmetric.  Warps
are also linked to large scale disk horizontal asymmetries, both
signatures showing that the outer spiral disks are not as well
virialized as the inner optical disks.  

\begin{figure}
\resizebox{\hsize}{!}{\includegraphics[]{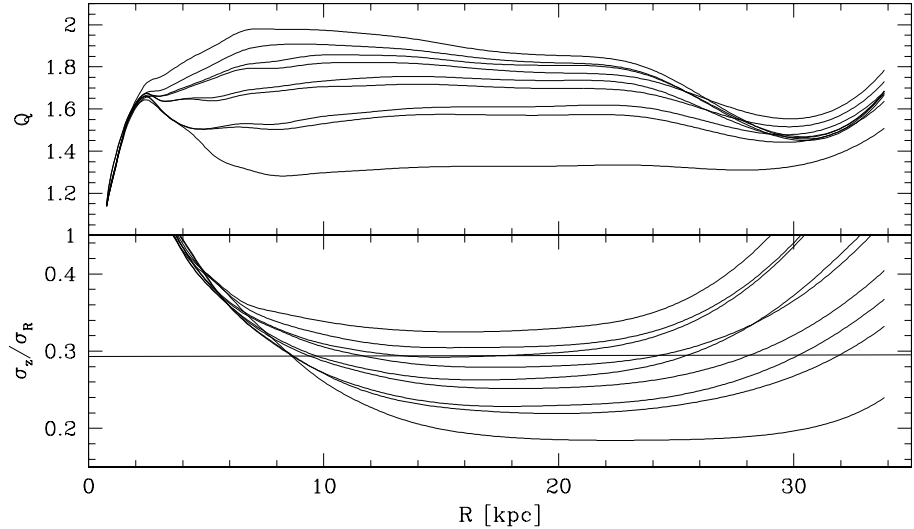}}
\caption{
  The radial stability parameter $Q$ (top) and the ratio
  $\sigma_z/\sigma_R$ (bottom) as a function of the galactic radius
  for different models.  In both graphics, the curves correspond, from
  bottom to top, in the interval $R = 10-20$, to the models of
  increasing thickness, respectively.}
\label{szrq}
\end{figure}

To answer the question about the general cause of warps, we have first
tried to well understand the dynamics of ideal isolated and purely
self-gravitating disks of collisionless particles.  In a second step
we will introduce energy dissipation, since disks form for the single
reason that the energy dissipation rate is much faster than the
angular momentum transport rate.  Therefore energy dissipation must be
taken as the second most important factor in understanding galaxies,
after the pure gravitational dynamics of collisionless matter.

Consequently we have undertaken first to study in detail massive
self-gravitating disks with various degrees of flattening by means of
N-body simulations (Revaz \& Pfenniger, 2004).  The simulated disks are
made of a stellar bulge and an exponential disk components, and a
collisionless heavy disk component proportional to the HI disk,
including a density depression in the optical disk, and a flaring
thickness almost proportional to radius.  The Milky Way is the
template galaxy for guiding the choice of the various mass ratios and
scale lengths. The mass components and profiles are also such that an
almost flat rotation curve are obtained for any thickness of the heavy
disk component.  By solving the Jeans equations separately for each
mass component, we can start simulations with an almost equilibrium
model, but with various degrees of velocity dispersion ratios
$\sigma_r/\sigma_z(R)$, while keeping an initial Toomre parameter $Q$
well above 1 on almost the full radial range (see Fig.~\ref{szrq}).

\begin{figure}
\resizebox{\hsize}{!}{\includegraphics[angle=-90]
    {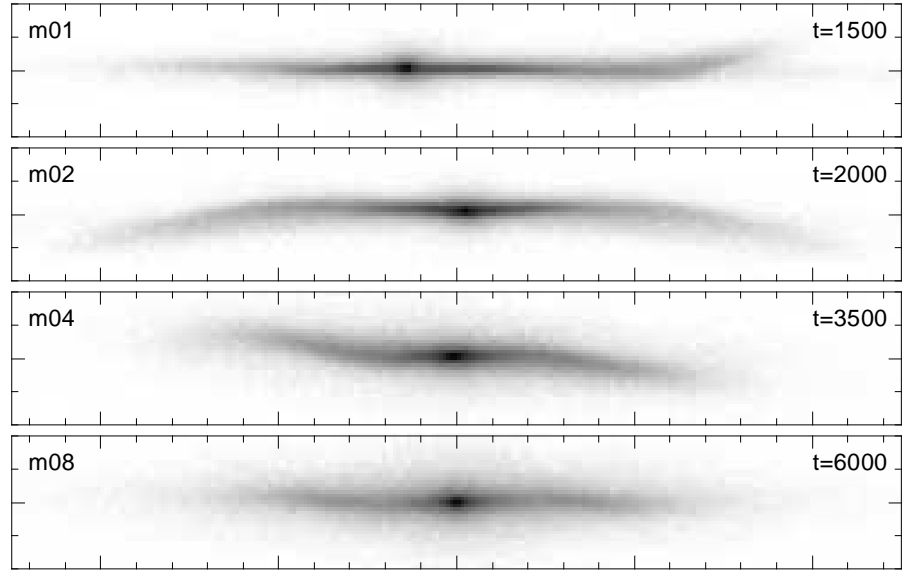}}
\caption{Edge-on projections of different heavy disk models having 
  developed spontaneously a warp.  The box dimensions are $100 \times
  16\,\rm{kpc}^2$}
\label{snapshots}
\end{figure}

The main result is that conformally to predictions made long ago by
Toomre (1966) and Araki (1985), too flat disks are unstable with
respect to bending instabilities.  The instability in thin sheets is
just related to the velocity dispersion ratios between the vertical
and radial velocity dispersions $\sigma_z$ and $\sigma_R$.  When
$\sigma_z/ \sigma_R < 0.293$ the sheet bends spontaneously with growth
rates of order of Gyr.  In thin disks this translates to first
S-shaped warp growing modes for slightly unstable disks, and secondly
for strongly unstable disks to U-shaped warp modes persisting for Gyrs
(see Fig.~\ref{snapshots}).

If, like bars and spiral arms, warps results of internal disk
instabilities, not only we obtain a unified picture of galaxies, but
also several new clues about the dark matter nature and its
distribution.  For the same fundamental cause, the marginal stability
of self-gravitating disks subject to an energy dissipation, disk
produce spontaneously bars, spirals and warps that counteract
dissipation by mechanical heating.

In order to be in such a warped state self-gravitating disks must first
obviously have a dissipative component.  Dusty gas can be identified as
the primary cause of energy losses.  Second, the mass distribution must
be sufficiently self-gravitating and thin in order to reach the Araki
stability threshold.  This provides an interesting constraint for the
gravitating mass.

The Araki criterion immediately tells us that a disk made of classical
smooth gas would never be transversally unstable because the gas
pressure would be isotropic.  But a classical gaseous disk dissipating
its heat decreases correspondingly its pressure, and inevitably after
some time reaches the Safronov-Toomre radial instability threshold, at
which point the subsequent non-linear evolution depends on the
detailed microscopic physics of the gas.  In the galaxy disk case we
know that the interstellar medium is widely non-homogeneous, which
means that the isotropic pressure assumption is not necessarily valid.
We also know well the most visible result of the gas instability, star
formation.  Once star form they evolve as collisionless matter, and
disks with such fluids are well known to evolve towards anisotropic
dispersions with $\sigma_z/ \sigma_R$ ratios of the order of 0.5
(Pfenniger \& Friedli, 1991; Huber \& Pfenniger, 2001).

Therefore the Araki criterion can only be met with a combination of
collisionless matter property to be able to have rather large velocity
dispersion anisotropy, and a dissipative component, which lowers over
time faster the velocity dispersion in $z$ than in $R$.  The dynamical
heating produced radially by spirals and bars increases also the
velocity anisotropy if the radial heating is inefficiently transferred
transversally to the plane.  The radial dynamics of disks indeed heats
effectively through bars and spiral arms essentially the radial
kinematics, maintaining it above $Q \sim 1$.

\section{The two types of dark matter }
So the ubiquitous existence of warps in disk galaxies is a strong hint
that they are sufficiently massive and thin to be transversally
unstable to warp modes, preferentially S-shaped modes. U-shaped modes
are also possible if a disk is driven sufficiently deep below Araki's
threshold.  If this is the case then we must have a substantial mass
component almost as thin as HI-disks that behaves as collisionless
matter for several rotational periods.  In order to regenerate warps,
dissipation is essential, without it a too anisotropic disk heats
dynamically until a stable thicker state is reached.

If disks react to bending instabilities by kinematical heating
transverse to the disk, then one must expect that warped disks are
maintained close to the marginal state balancing gas dissipation with
dynamical heating.  

Then the question is whether such warps may constrain the traditional
thick and hot dark halos made of collisionless matter.  By adding a
corresponding potentials to the initial N-body models, we have
calculated up to which halo mass with given flattening a marginal
state to bending would be kept (Revaz \& Pfenniger, 2004).  It turns
out that the effective disk thickness provides a strong constraint on
the dark halo mass, but almost no constraint on its flattening.  In
all studied cases the exact halo flattening is very little constrained
by the marginal stability state above a density flattening around
$0.3-0.5$, which is anyway the range usually considered in
cosmological simulations.  In contrast, the relative mass of a hot
dark halo within a radius comparable to the HI disk radius is directly
related to the precise massive disk thickness: the thicker the
marginally unstable disk is, the lesser mass can be contained in hot
spheroidal thick halo assuming that the warp results from a bending
instability.  For models parameters fitted to the Milky Way, the halo
is at most as heavy as the disk.

Since we can estimate the HI disk thickness, we can give a constraint
of the dark halo mass if the disk dark matter has a thickness similar
to the HI.  The Milky Way has a known HI thickness and a known warp,
therefore if this warp results from a disk marginal stability state
then the dark halo mass within a radius similar to the HI disk radius
($\sim 30\,\rm kpc$) is constrained to be below 0.4.  This value is
similar to the dark matter fraction found in evolved Sa's, S0's.
 
\section{Conclusions}

By realizing that bars, spirals and warps are different effects
resulting from that same fundamental causes, a slightly energy
dissipative gas component acting secularly on a mostly collisionless rotating
self-gravitating system, we obtain new clues about dark matter.  

First confirming several studies about the horizontal dynamics of
galaxies, we arrive at the conclusion that spiral disks are to first
order self-gravitating and collisionless, they must contain more matter
than seen.  This matter must be weakly collisional in order to develop
anisotropic velocity dispersions in $R$ and $z$.  It is clear that the
radial velocity dispersion is rapidly regulated by the radial
dynamical instabilities, so the anisotropy increases in $z$ through a
dissipative component, identified with the dusty gas.

Dissipation must be sufficiently effective in order to maintain the
disks close to instability, as witnessed by the spirals and warps, but
also most of the mass cannot be strongly collisional, otherwise the
velocity dispersion would be isotropic and no bending instability
would occur.  The physical solution that we are investigating is close
to the clumpuscule model in Pfenniger \& Combes (1994), where much of
the mass is condensed in the form of cold, dense planet-mass molecular
hydrogen clumps, stabilized in their core by a solid or liquid phase
of molecular hydrogen (Pfenniger, 2004ab; Revaz \& Pfenniger, 2004b).

The warped N-body models do not rule out traditional massive thick
halos with extended core.  However the mass of the massive halo can
hardly exceed the massive disk mass if this disk possesses a warp
produced by a bending instability.  The models do not constrain well
the dark halo axis ratio, provided it is above $\sim 0.3$.  

Thus we arrive at a bi-modal solution for dark matter in
galaxies that seems to satisfy all the known constraints, from
observational to galaxy dynamics and evolution constraints, and to
cosmological constraints.  Of course the nature of the dark matters
remain to be better understood and discovered by observations.
Encouragingly, tiny clumps of molecular hydrogen have been recently
detected (Heithausen, 2004), but the nature of the hot, non cuspy dark
halos remain to be found.

\begin{acknowledgments}
This work has been supported by the Swiss National Science Foundation.
\end{acknowledgments}


\end{document}